\documentclass[runningheads]{llncs}
\usepackage[T1]{fontenc}
\usepackage{graphicx}
\usepackage{xcolor}
\usepackage{amsmath,amssymb,amsfonts}
\usepackage{algorithmic}
\usepackage{graphicx}
\usepackage{textcomp}
\usepackage{algorithm}
\usepackage{hyperref}
\usepackage{listings}
\usepackage{comment}
\usepackage{hyperref}

\definecolor{codegreen}{rgb}{0,0.6,0}
\definecolor{codegray}{rgb}{0.5,0.5,0.5}
\definecolor{codepurple}{rgb}{0.58,0,0.82}
\definecolor{backcolour}{rgb}{0.95,0.95,0.92}

\lstdefinestyle{mystyle}{
    backgroundcolor=\color{backcolour},   
    commentstyle=\color{codegreen},
    keywordstyle=\color{magenta},
    numberstyle=\tiny\color{codegray},
    stringstyle=\color{codepurple},
    basicstyle=\ttfamily\footnotesize,
    breakatwhitespace=false,         
    breaklines=true,                 
    captionpos=b,                    
    keepspaces=true,                 
    numbers=left,                    
    numbersep=5pt,                  
    showspaces=false,                
    showstringspaces=false,
    showtabs=false,                  
    tabsize=2
}

\newcommand{\DOF}{\ensuremath{N_{\rm DOF}}}

\begin{document}
%
\title{A Sub-linear Low-Rank Solver for Poisson's Equation using Machine Learning Frameworks for GPU Acceleration}
\titlerunning{A Sub-linear Low-Rank Poisson Solver}
%
\author{M{\aa}ns I. Andersson\inst{1,2}\orcidID{0000-0002-6384-2630} \and
Daniel Appel{\"o} \inst{1}\orcidID{0000-0002-0378-4563}
}
\authorrunning{M. I. Andersson and D. Appel{\"o}}
%
\institute{Virginia Polytechnic Institute and State University (Virginia Tech),\\ Blacksburg, Virginia, United States of America \and
\email{mansande@vt.edu}
}
\maketitle              
\begin{abstract}
In this paper we explore a fast Poisson solver for problems with a solution that is known to be low-rank. We use an adaptive and warm started cross approximation called Cross-DEIM that iterates between index selection and and cross approximation to generate a low-rank solution. 
This paper focuses on leveraging a modern machine learning framework, PyTorch, as a general purpose array language to implement low-rank solvers based on Cross-DEIM. PyTorch enables native access to GPUs and accelerators but with a user-friendly high-level interface. We investigate statistical leverage scores for the index selection for the cross approximation due to the cost associated with the pivoted algorithms used with the discrete empirical interpolation methods (DEIM and QDEIM) which are historically preferred.
The cross approximation is naturally paired with a Discrete Sine Transform (DST) Poisson solver. This allows the Fast Fourier Transform (FFT) to be evaluated in batches along dimensions independently without any global transpose even in higher dimensions.
We present performance results running on a A100 GPU and AMD EPYC CPU demonstrating the usefulness of the approach that enables problems sizes that previously were not feasible. 

\end{abstract}
\section{Introduction}
Solving Partial Differential Equations (PDE) is important in many fields of science and engineering with applications such as computational fluid dynamics~\cite{lomax2001fundamentals} and computational electromagnetics~\cite{taflove2005computational}. Assuming a PDE has been discretized with $\DOF$ number of degrees of freedom the computational cost and memory usage is traditionally expected to scale, at best, as $\mathcal{O}(\DOF)$. For certain problems (e.g. elliptic PDE) the solution to the PDE may exhibit so called low-rank structure \cite{Bachmayr_2023}. When present, such low-rank structures makes it possible to design numerical approximation methods with exponentially better complexity, $\mathcal{O}(r^q (\DOF)^{\frac{1}{D}})$ in $D$ dimensions. Here $r$ is a measure of the rank of the solution (in this paper $r$ will be the matrix rank) and the parameter $q$ depends on the precise low-rank compression format used but is typically 2-4. 

In two dimensions a natural low-rank format is the singular value decomposition. In higher dimensions there are multiple competing formats for example Tucker tensors and Tensor Trains (TT) (see \cite{BallardKolda2025Tensor}). In this study we focus on the two dimensional case and low-rank techniques that can accelerate PDE discretization methods based on logically Cartesian meshes (e.g. finite difference methods and finite / spectral elements on quad elements). 

The methods we consider here are based on access to elements of a nonlinear function $G(i_1,i_2,\ldots,i_D)$ defined on a $D$-dimensional tensor. For example, $G$ may come from a spatial discretization of a PDE. As mentioned previously we specialize to the case $D=2$ and thus assume that $G$ is a matrix. Note that we allow that $G$ depend on another matrix $X$, i.e. an element of $G(X):$ $G(X)(i,j)$ can be a nonlinear function of elements $X(i^\prime,j^\prime)$. An example would be that $G$ is an approximation to $X_{xx} + X_{yy} + X^3$ using a local finite difference stencil, then $(i^\prime,j^\prime)$ would take values in, e.g., $(i \pm 1, i \pm 1)$. When $X$ is stored as a singular value decomposition with rank $r$ the elements in $X(i,j) = \sum_{k=1}^r \sigma_k u_k(i) v_k^T(j)$ can be accessed at $\mathcal{O}(r)$ cost.

One of the main challenges for efficient low-rank methods arises from the nonlinearity of $G$. For simplicity let the indices $i,j \in [1,n]$ so that the number of degrees of freedom of $G$ is $\DOF = n^2$. We are only interested in methods that scale sub-linearly in $\DOF$ and therefore any methods that require access to all of the elements of $G$ are excluded. Instead, to achieve sub-linear computational scaling, we use cross approximation (descried below, see also \cite{goreinov1997theory,goreinov2001maximal}) for computing low-rank approximations to terms like $G(X)$. Here we extend the cross approximation developed in \cite{appelo2025lraa}, called Cross-DEIM. 

As mentioned above, the original goal of the methods we study is to reduce computational complexity by removing computations that scale with $\DOF$ to instead scale as $\mathcal{O}(r^q (\DOF)^{\frac{1}{D}})$. While this clearly is, mathematically speaking, more efficient, the underlying algorithms are more involved and potentially less well suited for GPU acceleration. Here we study how new and existing solution methods, which exploit low-rank structures, can be adapted to GPUs using a high-level programming model. Specifically, in this paper we adapt algorithms that traditionally have been explored and developed in MATLAB or Julia to PyTorch. Our goal is not only to enable better utilization of modern hardware but also to provide an approachable interface for users, acknowledging that Python has become the de facto language for many parts of scientific computing and data analytics. It has become apparent with the rise of artificial intelligence  and machine learning that Python has a unique ability to glue together fast libraries and making them easy and productive to use. This has led to many Python based high-level frameworks and domain specific languages for ML that leverage GPUs, PyTorch~\cite{paszke2019pytorch}, TensorFlow~\cite{10.5555/3026877.3026899}, JAX~\cite{jax2018github}, Mojo~\cite{mojo}, etc. In this work we primarily use PyTorch, complemented with JAX implementations of functionality not available in PyTorch yet.

\section{Background}
We now discuss the mathematical and algorithmic details of the methods we accelerate.   
\subsection{Cross Approximation}
This section  describes  how to find a low-rank SVD approximation of a matrix $G \in \mathbb{R}^{m \times n}$ using Cross-DEIM, introduced in~\cite{appelo2025lraa}. The idea for the approximation is similar to SVD with truncation, in that it finds a low-rank approximation to a matrix but it is different than the SVD, in that, it achieves sub-linear scaling in the number of elements ($\DOF = mn$) of the matrix. This sub-linear scaling is possible since Cross-DEIM  only samples a few of the columns and rows of the matrix. Cross-DEIM is built from a combination of so called, cross-approximation, and index selection. We first discuss cross-approximation.

\begin{algorithm}[t]
\caption{{\tt [U,S,V,rC,rR] = scross(G,I,J)} \\Stabilized cross  approximation of $G$. \label{algo:scrossCUR}}
 \begin{algorithmic}[1]
 \STATE {\bf Input:}  $G$, and two index sets $\mathcal{I}$, $\mathcal{J}$ with $k$ and $l$ elements, respectively.
\STATE {\bf Output:} Approximate SVD of $G$, $U\in \mathbb{R}^{m \times r},$ $S \in \mathbb{R}^{r \times r}$, $V\in \mathbb{R}^{r \times n}$, and two vectors to test for linear dependence $r_{\rm R}, r_{\rm C} \in \mathbb{R}^r$.
\STATE $C = G(:,\mathcal{J}) \in \mathbb{R}^{m \times k}, \ \ R = G(\mathcal{I},:) \in \mathbb{R}^{l \times n}$  
\STATE $C = QR_{\rm C}$, \ \ $R^T = ZR_{\rm R}$ \COMMENT{Perform QR.}
\IF{$k \le l$}
\STATE Solve $Q(\mathcal{I},:) W = R$ \COMMENT{Solved using {\tt \textbackslash} if $Q(\mathcal{I},:)$ is well conditioned,}
\STATE \COMMENT{else solved by truncated SVD pseudoinverse.} 
\STATE $W = \hat{U}SV^T$  \COMMENT{Perform truncated SVD.}
\STATE $U = Q\hat{U}$
\ELSE
\STATE Solve $Z(:,\mathcal{J}) W = C^T$ \COMMENT{{\tt torch.linalg.lstsq} if $Z(:,\mathcal{J})$ is well conditioned,}
\STATE  \COMMENT{else solved by {\tt torch.linalg.pinv}.} 
\STATE $W^T = US\hat{V}^T$  \COMMENT{Perform truncated SVD.}
\STATE $V = Z\hat{V}$
\ENDIF
\STATE Return $USV^T \approx G$ and $r_{\rm R} = {\sf diag}(R_R), r_{\rm C} = {\sf diag} (R_C)$.
\end{algorithmic}
\end{algorithm}

Given a low-rank matrix $G \in \mathbb{R}^{m \times n}$ cross-approximation attempts to find a matrix $M \in \mathbb{R}^{r_1 \times r_2} $ such that 
\begin{align}
    G \approx G(:,\mathcal{J}) M G(\mathcal{I},:).
\end{align}
Here $\mathcal{I}$ and $\mathcal{J}$ are two index sets selecting $r_2$ rows and $r_1$ columns, respectively. We choose $M = G(\mathcal{I},\mathcal{J})^{+}$ where the pseudo-inverse can be computed in a stable manner using the techniques from \cite{donello2023oblique,appelo2025lraa}. 

Given a function handle to $G(i,j)$ and the two index sets $\mathcal{I}$ and $\mathcal{J}$ the subroutine {\tt scross}, detailed in Algorithm  \ref{algo:scrossCUR}, returns a singular value decomposition \mbox{$USV^T = G(:,\mathcal{J}) M G(\mathcal{I},:) \approx G$}.

Not surprisingly, the quality of the approximation $USV^T$ to $G$ depends heavily on the selected rows and columns in $\mathcal{I}$ and $\mathcal{J}$. Unfortunately, the solution to the optimal index selection problem (how to choose the $r_1, r_2$ best rows and columns) requires structural knowledge of $G$, e.g. its singular value decomposition. Fortunately, the Catch-22 situation, ($U \rightarrow \mathcal{I} \rightarrow U \rightarrow \cdots$) can be resolved by iteration, and this is the strategy used in Cross-DEIM.  

We note that the cross approximation is distinct from the (often denoted CUR) approximation that computes $M$ by projection $M = G(:,\mathcal{J})^+GG(\mathcal{I},:)^+$. The CUR approximation requires access to the entire matrix $G$ it cannot be computed with sub-linear cost in $\DOF$.

\subsection{Index Selection}
Before describing the Cross-DEIM method we first describe the index selection methods we use here.  

\begin{algorithm}[t]
\caption{{\tt [$\mathcal{I}$] = (Q)DEIM(U)} \\ Index selection using DEIM and QDEIM \label{algo:deim}}
 \begin{algorithmic}[1]
 \STATE {\bf Input:}  Orthogonal matrix $U$ of size $k \times l$, Index range $I$
\STATE {\bf Output:} Index set $\mathcal{I}$ of size $l$
\IF{DEIM}
\STATE  $[\sim,\sim, P]=\textrm{lu}(U,\textrm{'vector'})$ \COMMENT{Perform pivoted LU on $U$.}
\STATE $\mathcal{I}=P^TI$
\STATE Return $\mathcal{I}(1:l)$
\ELSIF{QDEIM}
\STATE  $[\sim,\sim, p]=\textrm{qr} (U^T,\textrm{'vector'})$ \COMMENT{Perform column pivoted QR on $U^T$.}
\STATE $\mathcal{I} = p$
\STATE Return $\mathcal{I}(1:l)$
\ENDIF
\end{algorithmic}
\end{algorithm}

\subsubsection{Discrete Empirical Interpolation Methods}
The discrete empirical interpolation method (DEIM)~\cite{doi:10.1137/090766498} and the QR based variant, QDEIM, \cite{doi:10.1137/15M1019271} can be used for index selection. QDEIM and DEIM (described in Algorithm~\ref{algo:deim}) have different strengths. When used together with CUR, QDEIM provides sharper error estimates than DEIM but the complexity of computing DEIM (via pivoted LU) has a slightly lower constant (LU is slightly cheaper than QR). Also, the indices produced in the DEIM algorithm are ordered by decreasing importance, thus it is possible to stop the process early (we don't explore this here) for some computational savings. 

\begin{algorithm}[t]
\caption{{\tt [$\mathcal{I}$] = LeverageScores(U)} \\ Index selection}
 \begin{algorithmic}[1]
 \STATE {\bf Input:}  Orthogonal matrix $U$ of size $k \times l$
\STATE {\bf Output:} Index set $\mathcal{I}$ of size $l$
\STATE $\pi_i = \frac{1}{k}\Sigma_{\eta=1}^{k} (u_i^{\eta})^2 $ 
\STATE Use $c = \mathcal{O}(k \log k/\epsilon^2) $
\STATE Keep $i^\text{th}$-column with probability of $p_j=\min\{1,c\pi_j\} $
\STATE Return $\mathcal{I}$
\end{algorithmic}
\label{algo:ls}
\end{algorithm}

\subsubsection{Statistical Leverage Scores}
In previous non-GPU implementations of Cross-DEIM QDEIM was the preferred algorithm for index-selection. Here we also consider statistical leverage scores (LS) described in~\cite{doi:10.1073/pnas.0803205106} and provided as an algorithm in Algorithm~\ref{algo:ls}. This is an algorithm mainly used in data analytics. While LS typically provides a worse index selection than DEIM based selection it is faster. Unlike DEIM and QDIEM, it does not rely on a pivoted factorization and therefore does not scale quadratically with the size of the index set. We also belive that the lower quality of the index selection by LS is less critical in an algorithm such as Cross-DEIM, where  the final index set is built up incrementally and its quality is checked adaptively. We perform numerical experiments  to understand when is possible if a method that needs more iterations to build up the Cross-DEIM index set can still be faster if each iteration is less costly.  

\begin{algorithm*}[]
  \caption{{\tt [U,S,V] = Cross-DEIM(G, U0, V0, $\epsilon$, $r_{\rm max}$, {\tt maxiter})} \\
  Adaptive Cross-DEIM approximation to  $G \in \mathbb{R}^{m \times n}$}
  \begin{algorithmic}[1]
  \STATE {\bf Input:} Matrix $G \in \mathbb{R}^{m \times n}$, initial rank $r$ guess to the singular vector matrix $U_0 \in \mathbb{R}^{m \times r}, V_0 \in \mathbb{R}^{n \times r}$, tolerance $\epsilon$, maximum output rank $r_{\rm max}$, maximum number of iterations {\tt maxiter}. 
  \STATE {\bf Output:} Approximate SVD of $G$, $U\in \mathbb{R}^{m \times r},$ $S \in \mathbb{R}^{r \times r}$, $V\in \mathbb{R}^{r \times n}$.
  \STATE Set $\mathcal{I}_0=\mathcal{J}_0 = \emptyset$.
\FOR{$k = 0, 1, \ldots$, {\tt maxiter}}
\STATE $\mathcal{I}_{k}^\ast = {\tt QDEIM}(U_{k-1})$, \ $\mathcal{J}_{k}^\ast = {\tt QDEIM}(V_{k-1})$ \COMMENT{ {\tt QDEIM} can be replaced by {\tt DEIM} or {\tt LeverageScores}}
\STATE $\mathcal{I}_{k} = \mathcal{I}_{k}^\ast \cup \mathcal{I}_{k-1}, \mathcal{J}_{k} \leftarrow \mathcal{J}_{k}^\ast \cup \mathcal{J}_{k-1}$ \COMMENT{Note: the index sets are ordered by {\tt (Q)DEIM} but not with {\tt LeverageScores}.}
\IF[Make sure that that the index set increase by at least one when using {\tt (Q)DEIM}]{\textcolor{blue}{$|\mathcal{I}_{k}| = |\mathcal{I}_{k-1}|$ or $k=0$}} 
\STATE \textcolor{blue}{$\mathcal{I}_{k} = \mathcal{I}_{k}^\ast \cup \{ i_{\rm rand} \in \complement (\mathcal{I}_{k}^\ast) \}$} \COMMENT{using a random $i_{\rm rand}$ from the complement of $\mathcal{I}_{k}^\ast$.}
\ENDIF
\IF{\textcolor{blue}{$|\mathcal{J}_{k}| = |\mathcal{J}_{k-1}|$ or $k=0$}} 
\STATE \textcolor{blue}{$\mathcal{J}_{k} \leftarrow \mathcal{J}_{k}^\ast \cup \{ j_{\rm rand} \in \complement (\mathcal{J}_{k}^\ast) \}$ }
\ENDIF
\STATE $[U_k,S_k,V_k,r_{\rm C},r_{\rm R}] = {\tt scross}(G, \mathcal{I}_k,\mathcal{J}_k)$ \COMMENT{Returns SVD approximation and $r_C$, $r_R$}
\FOR{$l = 0, 1, \ldots, |\mathcal{I}_k|$}
\IF{$|(r_{\rm R})_{l}| < 10^{-12}$}
\STATE Remove element $l$ from $\mathcal{I}_k$  \COMMENT{Remove redundant rows in $R = G(\mathcal{I}_k,:)$.}
\ENDIF  
\ENDFOR 
\FOR{$l = 0, 1, \ldots, |\mathcal{I}_k|$}
\IF{$|(r_{\rm C})_{l}| < 10^{-12}$}
\STATE Remove element $l$ from $\mathcal{J}_k$ \COMMENT{Remove redundant columns in $C = G(:,\mathcal{J}_k)$.} 
\ENDIF
\ENDFOR 
\STATE $\rho = \| U_{k}S_{k}V_{k}^T- U_{k-1}S_{k-1}V_{k-1}^T \|, \ \ S_{\rm min} =  \min({\sf diag}(S_k))$
\STATE  $\eta_1 = \| (I(:,\mathcal{I}_k))^T U_k \|_2^{-1}, \ \ \eta_2 = \| V^T_{k} I(\mathcal{J}_k,:) \|_2^{-1}$ 
\IF{$\max(\rho,\min(\eta_1(1+\eta_2),\eta_2(1+\eta_1))S_{\rm min})  < \epsilon$} 
\STATE Break out of for loop  \COMMENT{Above $S_{\rm min}$ is the smallest s.v. in the $k^\text{th}$ approx.}
\ENDIF
\ENDFOR
\STATE Find $r^\ast$ so that $\sum_{l = r^\ast+1}^{\min(m,n)} S_l^2 < \epsilon^2$ 
\STATE Set $r = \max(\min(r^\ast,r_{\rm max}),1)$
\STATE Return $U_k(:,0:r), S_k(0:r,0:r), V_k(:,0:r)$
\end{algorithmic}
\label{algo:crossDEIM}
\end{algorithm*}

\subsubsection{The Cross-DEIM algorithm}
Cross-DEIM iterates between; a) an approximate singular value decomposition for index selection and, b) cross approximation to generate a new low-rank approximation. We now describe the main steps in Algorithm~\ref{algo:crossDEIM}. The algorithm is initialized with empty row and column index sets and an initial guess of the approximate left and right leading singular vector matrices. Then at each iterate, given the current row and column index sets $\mathcal{I}_k, \mathcal{J}_k$ and approximate left and right leading singular vector matrix $U_{k-1}, V_{k-1},$ the iterative procedure updates the index sets and the SVD until it converges. 

Specifically, the index update is done by enriching the current index sets with indices selected by QDEIM with singular vector matrices $U_{k-1}, V_{k-1}$. As a safeguard, we add one additional randomly chosen index if no new indices are added in index selection step. Note that when LS is used its probabilistic nature makes it exceedingly unlikely that exactly the  same index set is selected twice in a row, and it is therefore possible to remove the logic on lines 7-12 (highlighted in blue). The stabilized cross approximation updates the singular vectors. We remove the redundant (linearly dependent) rows and columns as needed. The iteration is stopped when both the difference of the consecutive updates and the low-rank indicators are less than the provided tolerance.

\section{A Fast Low-Rank Poisson Solver}
We now demonstrate how Cross-DEIM can be leveraged to design a solver for the Poisson equation with sub-linear complexity in the number of degrees of freedom. To this end consider Poisson's equation on the unit square, $\Omega = [0,1]^2$ with homogeneous Dirichlet boundary conditions
\begin{equation}\label{eq:poisson}
    -\Delta u = f, \quad (x,y) \in \Omega, \ \ \ \  u = 0,  \quad (x,y) \in \partial\Omega.
\end{equation}
Let $W_{i,j}$ be a grid function on a two dimensional equidistant grid with grid spacing $h=1/(n+1)$ and with $n$ interior gridpoints in each dimension. Let the $n\times n$ matrix $T$ be tri-diagonal with elements $(-1,2,-1)$ on the diagonals. Then the approximation to (\ref{eq:poisson}) can be written as the matrix equation 
\begin{equation} \label{eq:matrix_poisson}
    TW + WT^T = h^2F,
\end{equation}
where $F$ is the matrix representing the grid function corresponding to the forcing $f$. The matrix $T = Z\Lambda Z^T$ has eigenpairs $(\lambda_k,z_k)$ whose expressions are
\[
\lambda_k = 2(1-\cos\left(\frac{\pi k}{n+1}\right)), \ \ z_k(i) = \sqrt{\frac{2}{n+1}} \sin\left(\frac{ik\pi}{n+1} \right).
\]
Therefore, the multiplication with the eigenvector matrix and a vector can be carried out in $\mathcal{O}(n \log n)$ operations.  

\subsection{Full-Rank Solver}
A classic approach (see e.g. \cite{demmel}) for solving \eqref{eq:matrix_poisson} is to multiply from the left with $Z^T$ and from the right with $Z$. Introducing $\hat{F} = Z^TFZ$ and $\hat{W} = Z^TWZ$ \eqref{eq:matrix_poisson} can be written 
\[
 \Lambda \hat{W} + \hat{W} \Lambda = h^2\hat{F},
\]
and solved element by element via the formula 
\begin{equation}\label{eq:g_for_p}
\hat{W}(i,j) = \frac{h^2 \hat{F}(i,j)}{\lambda_i + \lambda_j}.  
\end{equation}
Then, using the DST when computing $\hat{F}$ and $W$ from $\hat{W}$ costs $\mathcal{O}(n^2 \log n)$ operations and solving \eqref{eq:g_for_p} requires $\mathcal{O}(n^2)$ operations for a total of
\[ 
\mathcal{O}(\DOF \log( \DOF^{1/2})),
\]
operations. 
\subsection{Low-Rank Solver}
If \mbox{$F  = U_F S_F V_F^T$} is of rank $r$ and represented via its SVD factors we can compute  
\[
\hat{F} = \textsf{DST}(U_F) S_F ( \textsf{DST}(V_F))^T,
\]
at a reduced cost  $\mathcal{O}(r n \log n)$. 

Then, instead of solving \eqref{eq:g_for_p} for each element (which would require storage and compute $\mathcal{O}(n^2)$) we directly apply Cross-DEIM on \eqref{eq:g_for_p} to find a low rank representation to $\hat{W} = \hat{U} S \hat{V}^T$. When this solution is of $\sim r$ we expect that the Cross-DEIM solve can be done in $\mathcal{O}(r^2 n + r^3)$ operations. Recovering the SVD factors of $W$ amounts to computeing (assuming that the DST is normalized such that $\textsf{DST}(\textsf{DST}(I))=I$) 
\[
W = \textsf{DST}(\hat{U}) S ( \textsf{DST}(\hat{V}))^T,
\]
using  $\mathcal{O}(r n \log n)$ operations. 

In summary, when the solution and right hand side is of rank $\sim r < \sqrt{n}$ the dominant parts of the total cost for solving \eqref{eq:matrix_poisson} in terms of $\DOF = n^2$ is 
\begin{equation}
\label{eq:On12}
\boxed{\mathcal{O}\left(\DOF^{\frac{1}{2}} \left[ r^2 +  r \log  \DOF^{\frac{1}{2}} \right] \right).}
\end{equation}




\section{Methodology}
We use the PyTorch library~\cite{paszke2017automatic} to easily parallelize the code on diverse hardware. PyTorch does not support a pivoted QR required by QDEIM. 
Due to the large QR and SVD factorizations we link the MAGMA library~\cite{abdelfattah2024magma} through the \texttt{jax} library~\cite{jax2018github} and PyTorch, although the most specialized routines are not available.

\subsection{PyTorch and Just-in-time Compilation}

We utilize the PyTorch compiler (\texttt{@torch.compile}), which consists primarily of two technologies: 1) TorchDynamo, which captures Python-level execution into computational graphs that serve as an intermediate representation (IR), and 2) TorchInductor, which compiles these graphs into optimized low-level code for multiple hardware backends.

The compiler captures large regions of computation and lowers them into optimized kernels that can be executed more efficiently than standard eager-mode execution. By compiling entire graph regions together, PyTorch reduces Python interpreter overhead, framework dispatch costs, and GPU kernel launch overhead.

TorchInductor further improves performance through kernel fusion, where multiple operations are merged into a single kernel. This reduces memory traffic, minimizes intermediate tensor materialization, and lowers launch overhead. Additional optimizations are obtained through the use of specialized kernels for common operations such as GEMMs, Triton-generated GPU kernels, and CUDA Graphs for reducing CPU-side launch latency and improving execution efficiency on repeated workloads and especially small computations. 

The computation of the QDEIM is decorated with \texttt{@torch.compiler.disable} since functions from Jax are not native to PyTorch and cannot be correctly parsed into the graph and forces graph-breaks. Some important computations in the current formulation of Cross-DEIM requires control-flow based on tensors sizes in multiple places which creates graph-breaks. These breaks can limit the optimization opportunities. 

We fix the random number generator seed and perform warm-up runs using the same seed as the timed executions, ensuring that any JIT compilation and optimization overhead occurs prior to measurement. This allows the reported timings to reflect algorithm performance rather than compilation costs.

\subsection{Test Problems} 
Before presenting the real-world benchmark examples we show the correctness and performance behavior of the our modified method using well-understood test matrices. The following matrices are used for the analysis of the approximation ability of the implementations: 
\begin{align}
H(i,j) = \frac{1}{1 + i + j}, \quad \quad K(i,j) = \left(\frac{|x_i+y_j|}{2}\right)^5.
\end{align}
Here $H$ is the Hilbert matrix which is known to have rapid singular value decays. The matrix $K$ uses the grid $x_i = -1 + 2 \frac{i-1}{m-1}$ and $y_i = -1 + 2 \frac{j-1}{n-1}$. It is a $C^4$ function with a non-smooth feature along $x + y = 0$. The limited smoothness gives the  matrix $K$ slow singular value decay, which  is numerically challenging for cross approximation.

The fast Poisson solver is tested on a uniform two-dimensional grid with homogeneous Dirichlet boundary 
conditions with a right-hand-side that is constructed to generate a low-rank solution. 

\begin{equation}    
    f(x,y) = e^{-36((x-0.1)^2-(y-0.55)^2)}, 
\end{equation}
for $x, y \in [-1,1]$. We investigate grids of the size $n = 2^{10}$ to $n = 2^{18}$.

\subsection{Hardware and Software Setup}
The PyTorch implementation\footnote{\href{https://github.com/MaansAndersson/torchlr}{https://github.com/MaansAndersson/torchlr}} is evaluated  on two machines: one machine with a NVIDIA A100 GPU with 40 GB memory and the other a 48 core AMD AMD EPYC 9454 CPU located at Advanced Research Computing (ARC) at Virginia Tech.

The index selection is non-deterministic for all index selection methods; therefore, for the wall-clock-time evaluation of the whole application we run with ten different seeds. 

In \autoref{tab:libs} the main collection of the computational libraries used. The bulk of the code is implemented with PyTorch. We use Jax to implement the QDEIM algorithm as PyTorch does not provide a pivoted QR algorithm. We link with both \texttt{cuSOLVE} and \texttt{MAGMA} and let the framework choose the backend using its own heuristics.

\begin{table}[h!]
    \centering
    \caption{Libraries and versions used. Note that not all linear algebra routines from the \texttt{BLAS}/\texttt{LAPACK} implementations are available through PyTorch.}
    \begin{tabular}{lll}
      \textbf{Library}   & \textbf{Version} & \textbf{Usage}\\
      \hline \texttt{PyTorch}   & 2.11.0+cu130 & General array manipulation and library calls\\
      \texttt{Jax}       & 0.9.2 & General array manipulation and library calls\\
      \texttt{cuSOLVE}   & 13.0 & Linear Algebra Routines for GPUs \\
      \texttt{MAGMA}     & 2.6.1 & Linear Algebra Routines for GPUs \\
      \texttt{MKL}   &  2024.2 & Linear Algebra Routines for CPUs \\
      \texttt{GCC}   & 13.3 & C++ compiler 
    \end{tabular}
    \label{tab:libs}
\end{table}
\section{Results}
\begin{figure*}[h!]
    \centering
    \includegraphics[width=0.33\linewidth]{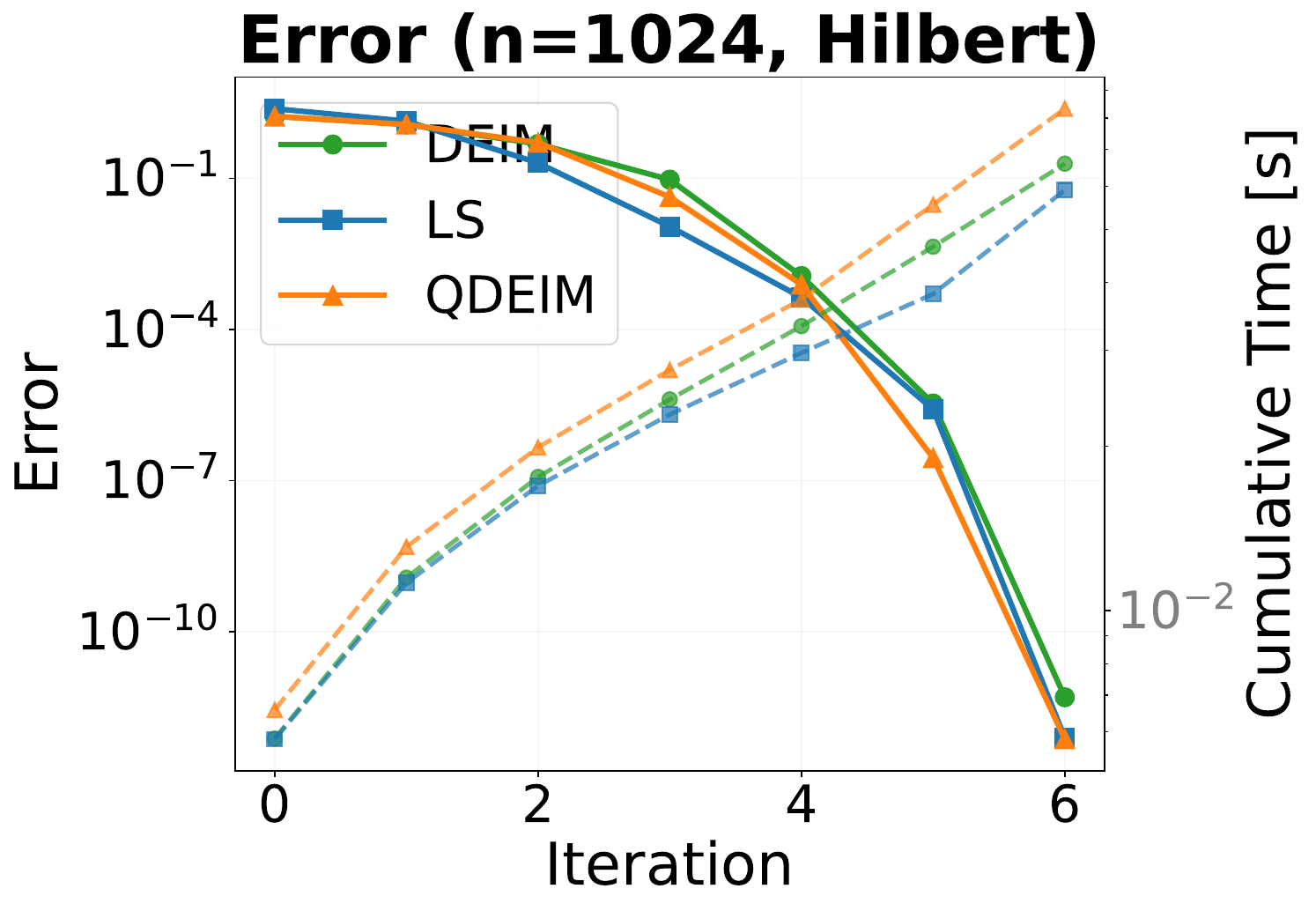} %
    \includegraphics[width=0.33\linewidth]{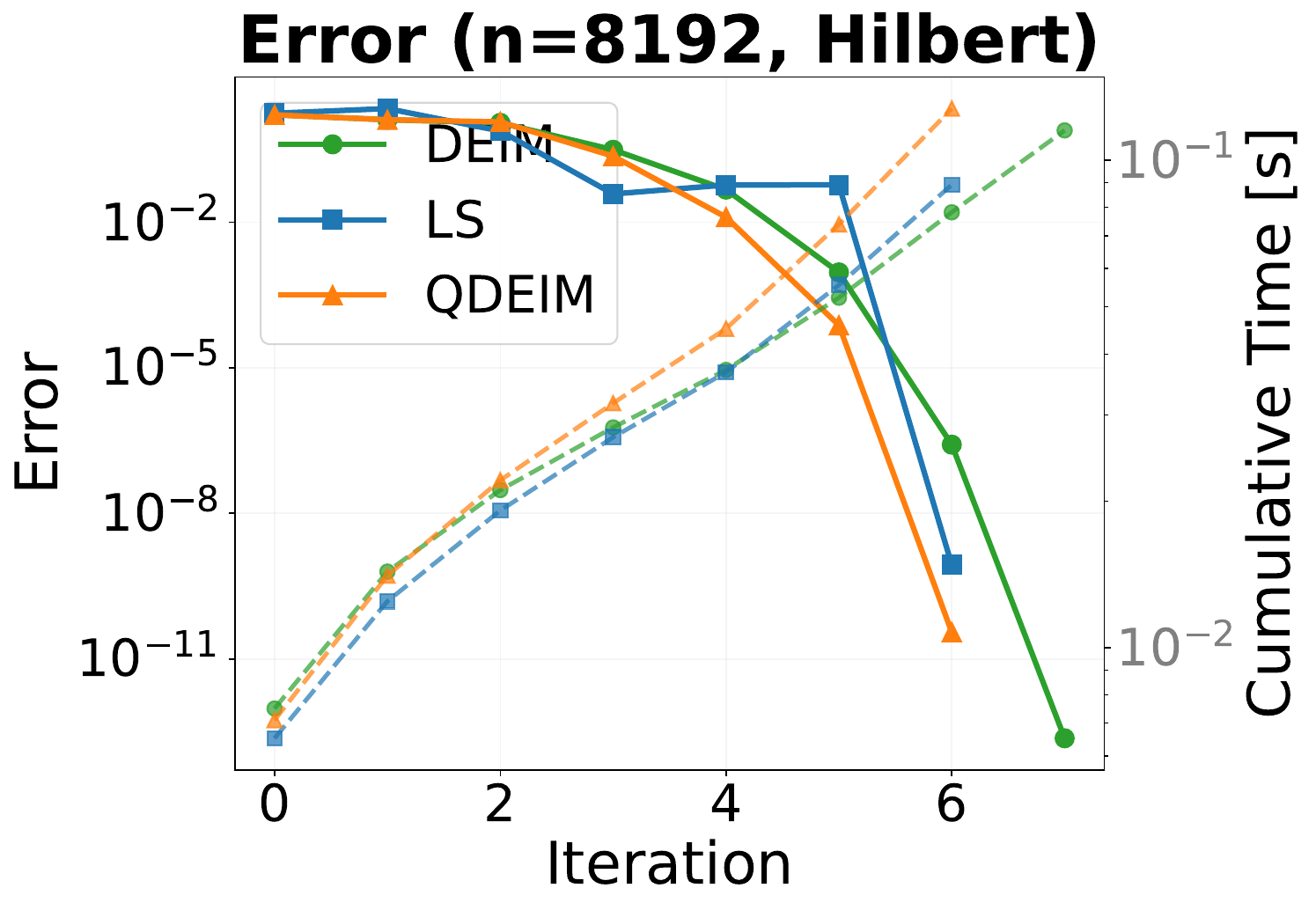}
    \includegraphics[width=0.32\linewidth]{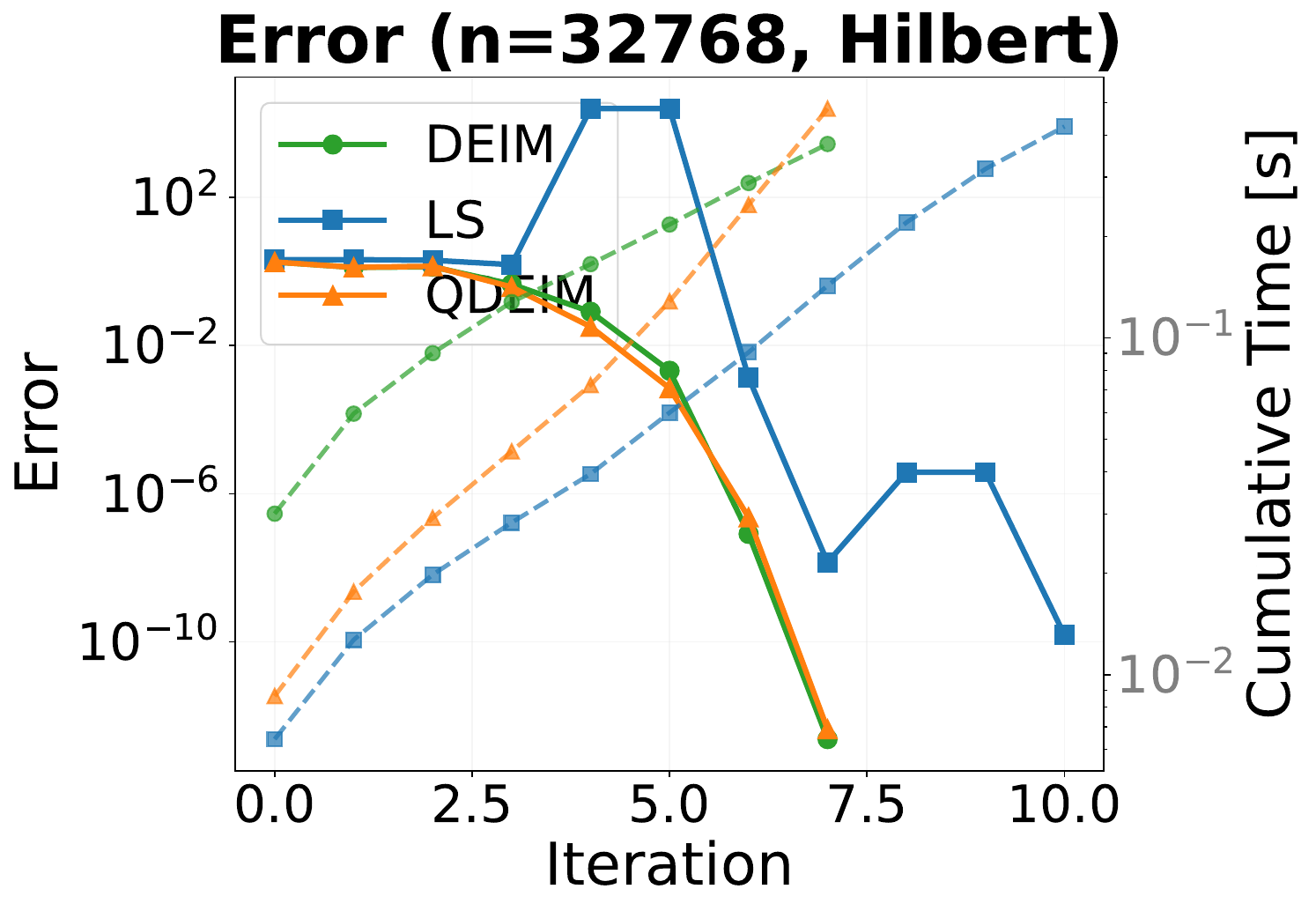}
    \includegraphics[width=0.33\linewidth]{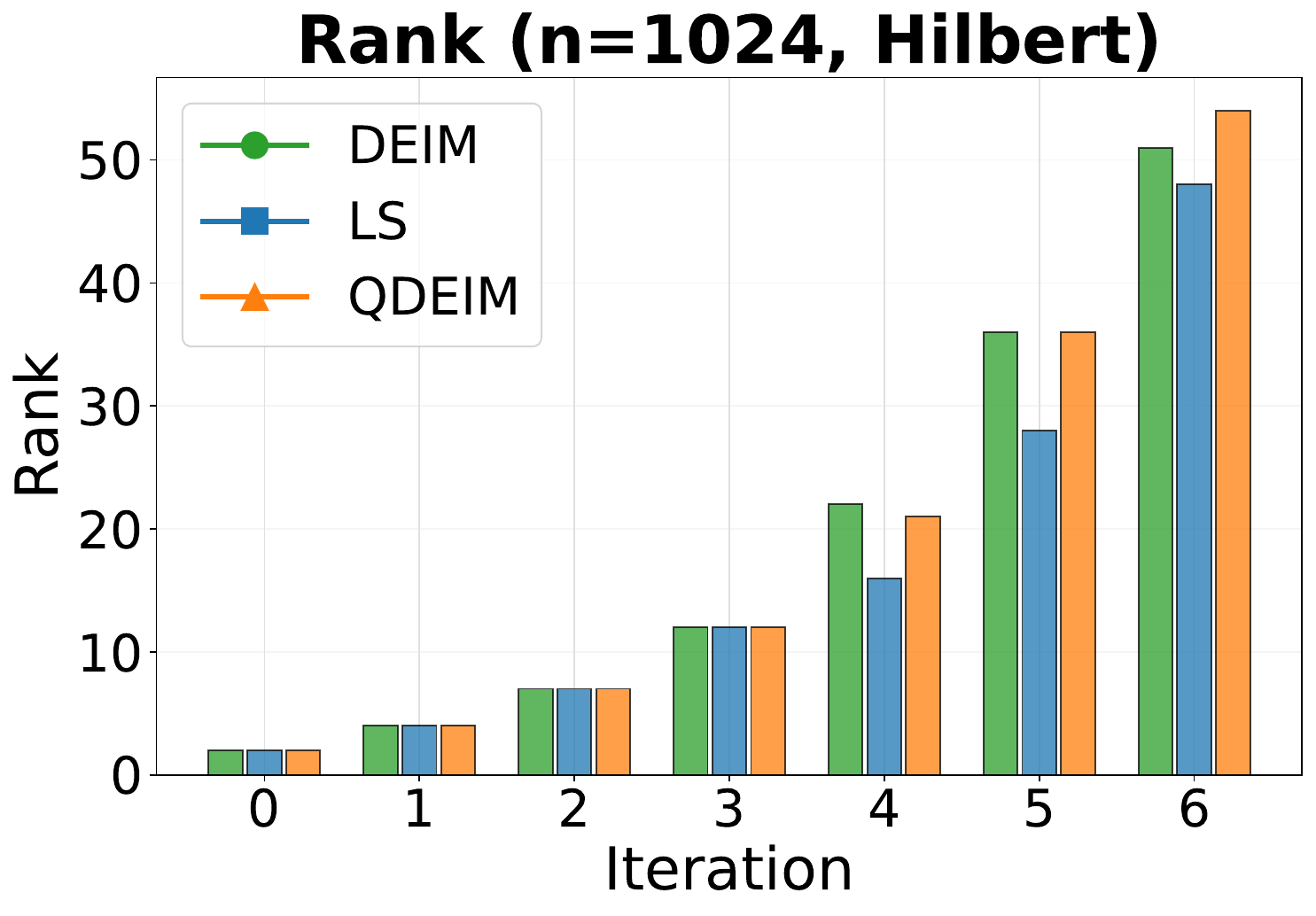}
    \includegraphics[width=0.33\linewidth]{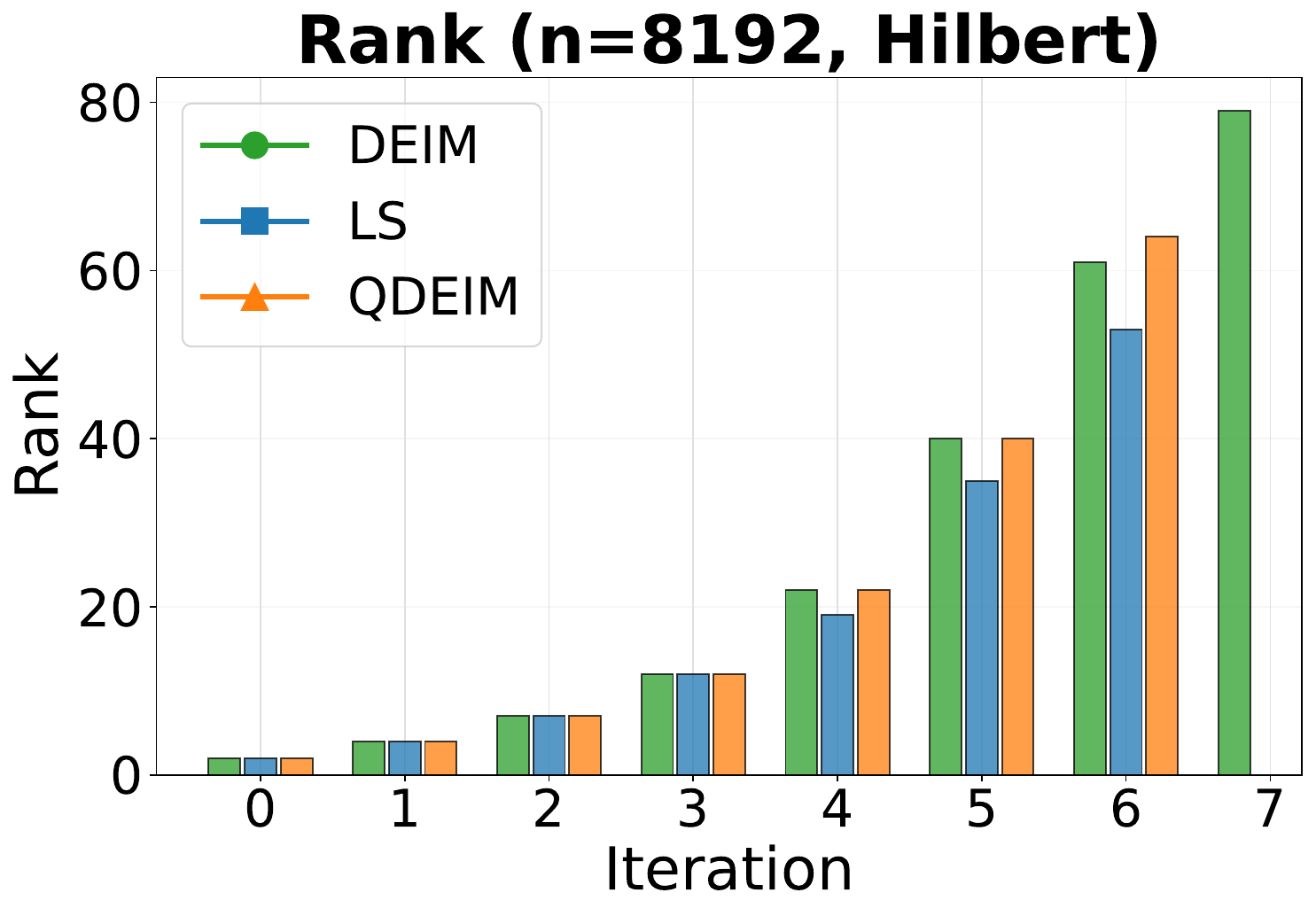}
    \includegraphics[width=0.32\linewidth]{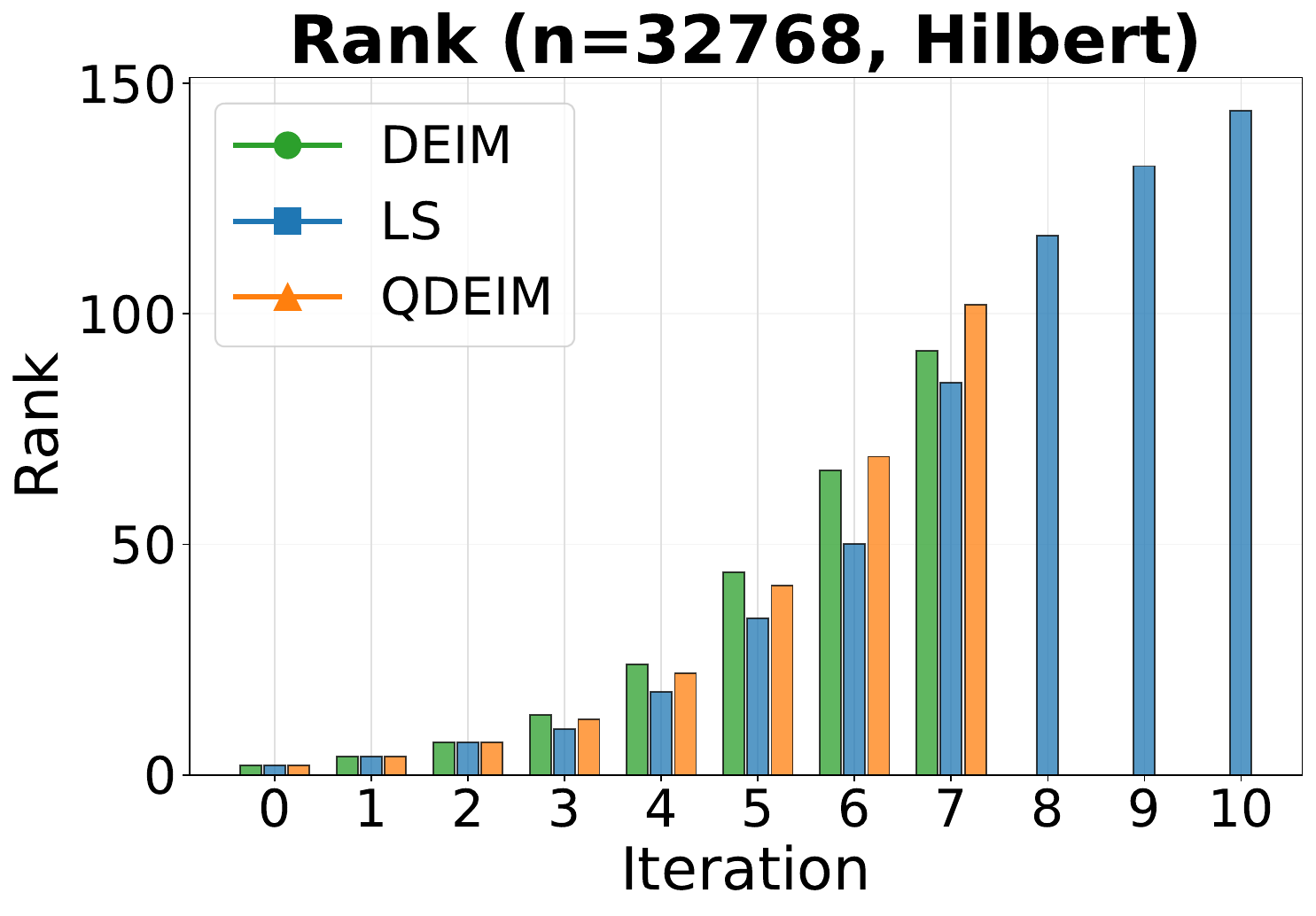}
    \caption{The top row presents the error (solid) and the cumulative time (dashed) for the Cross-DEIM implementation for three sizes of the Hilbert matrix. The second row presents the final rank of the solution at each iteration of Cross-DEIM.}
    \label{fig:indexselection}
\end{figure*}

\begin{figure*}[h!]
    \centering
    \includegraphics[width=0.33\linewidth]{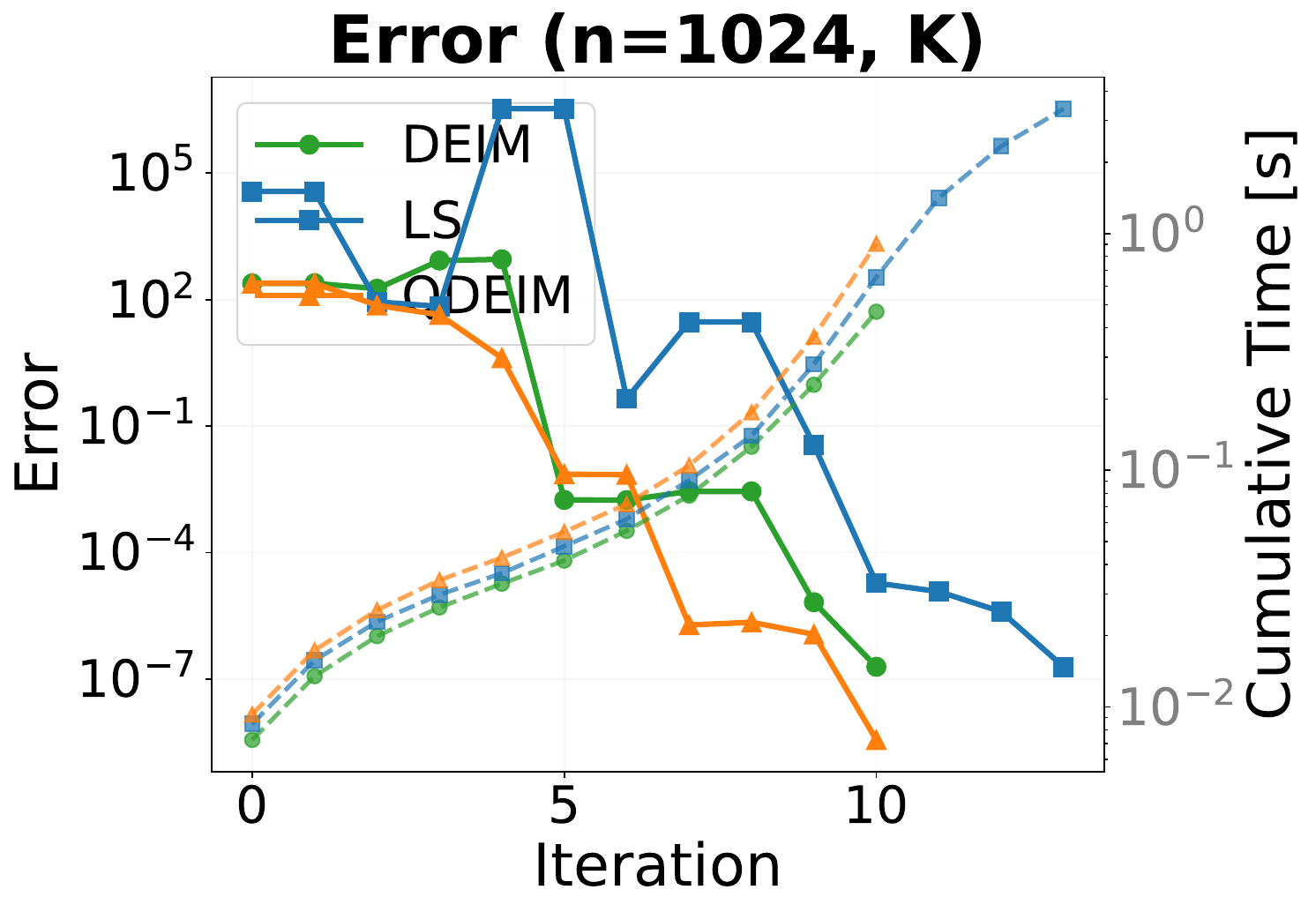} %
    \includegraphics[width=0.33\linewidth]{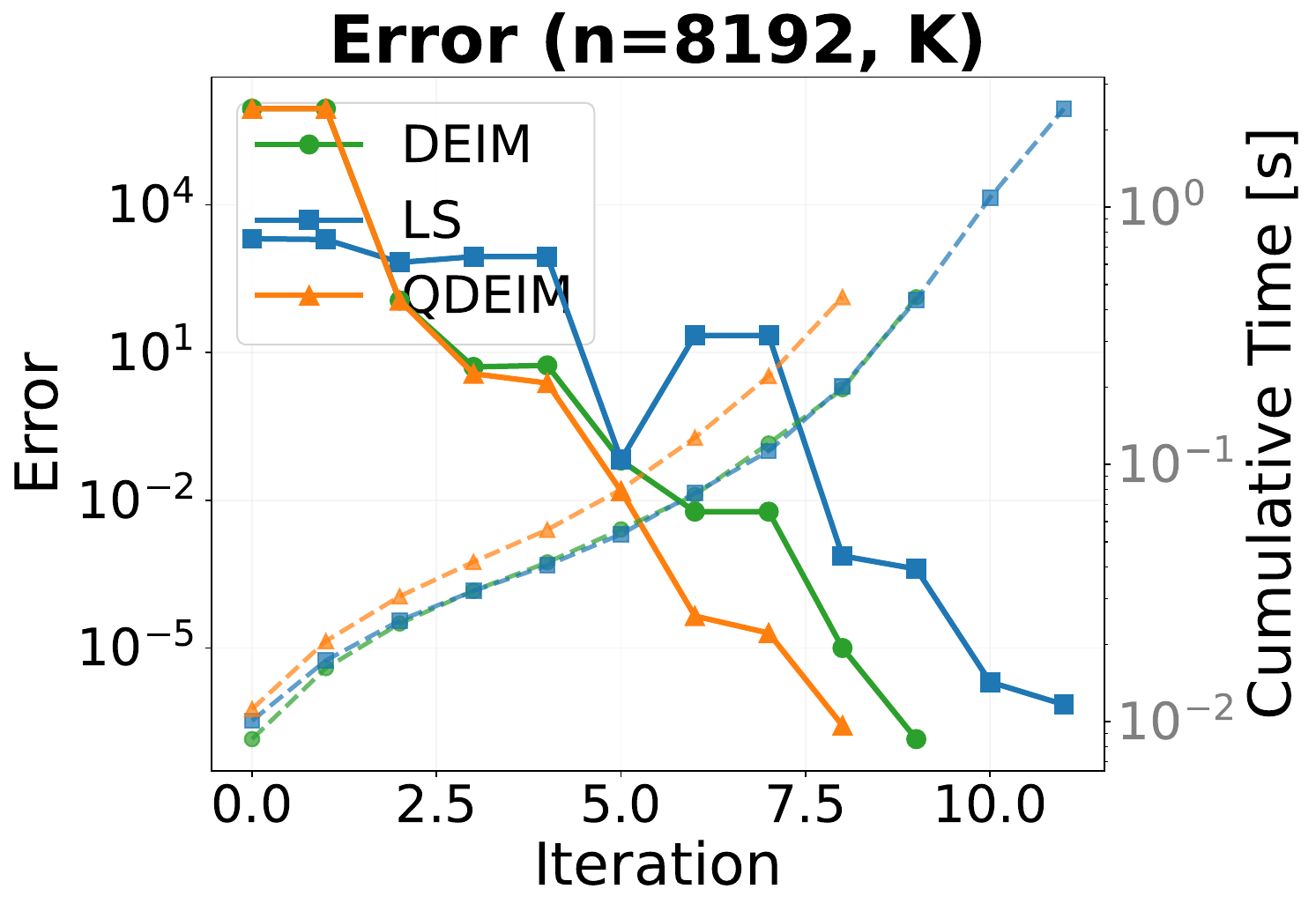}
    \includegraphics[width=0.32\linewidth]{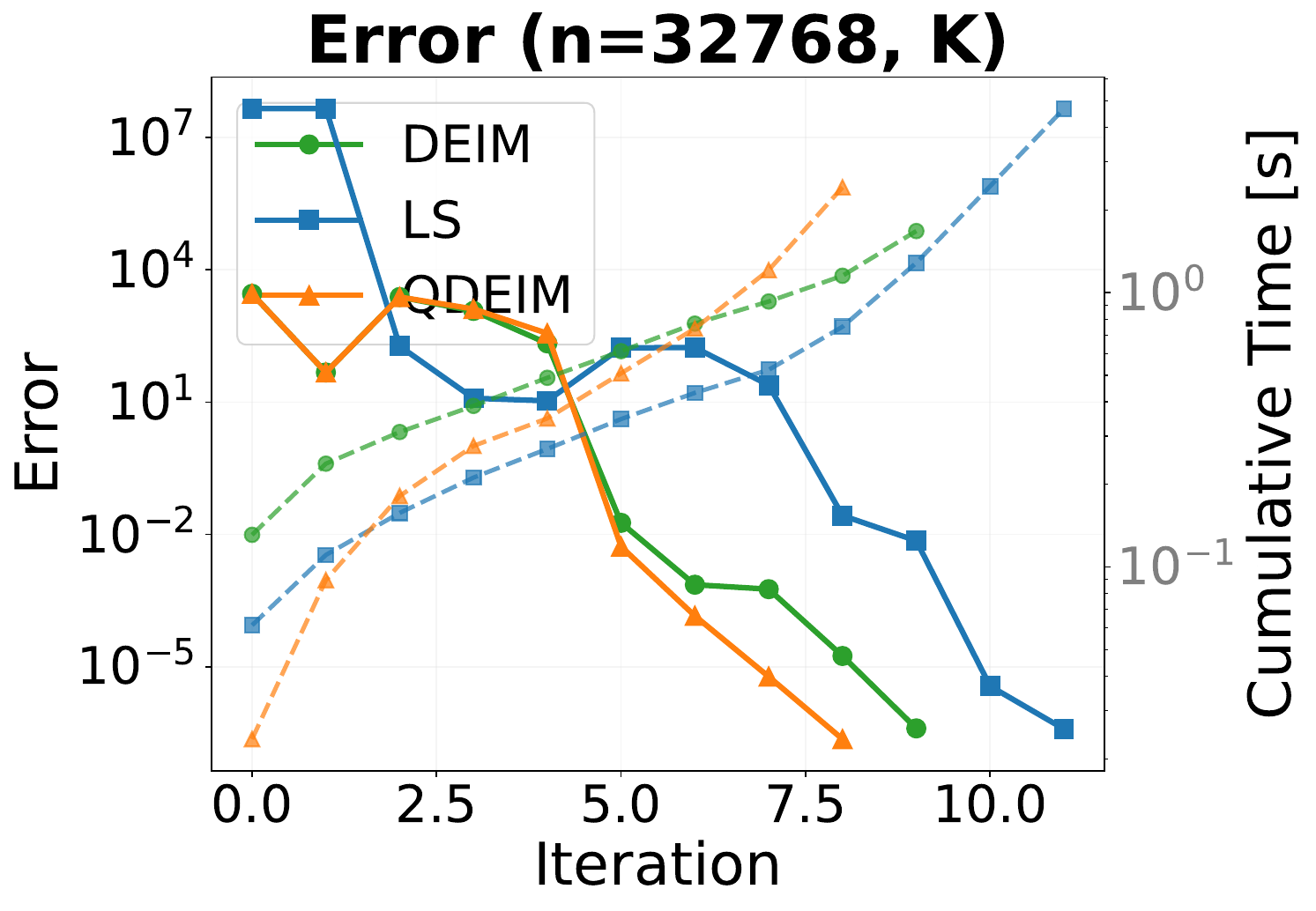}
    \includegraphics[width=0.33\linewidth]{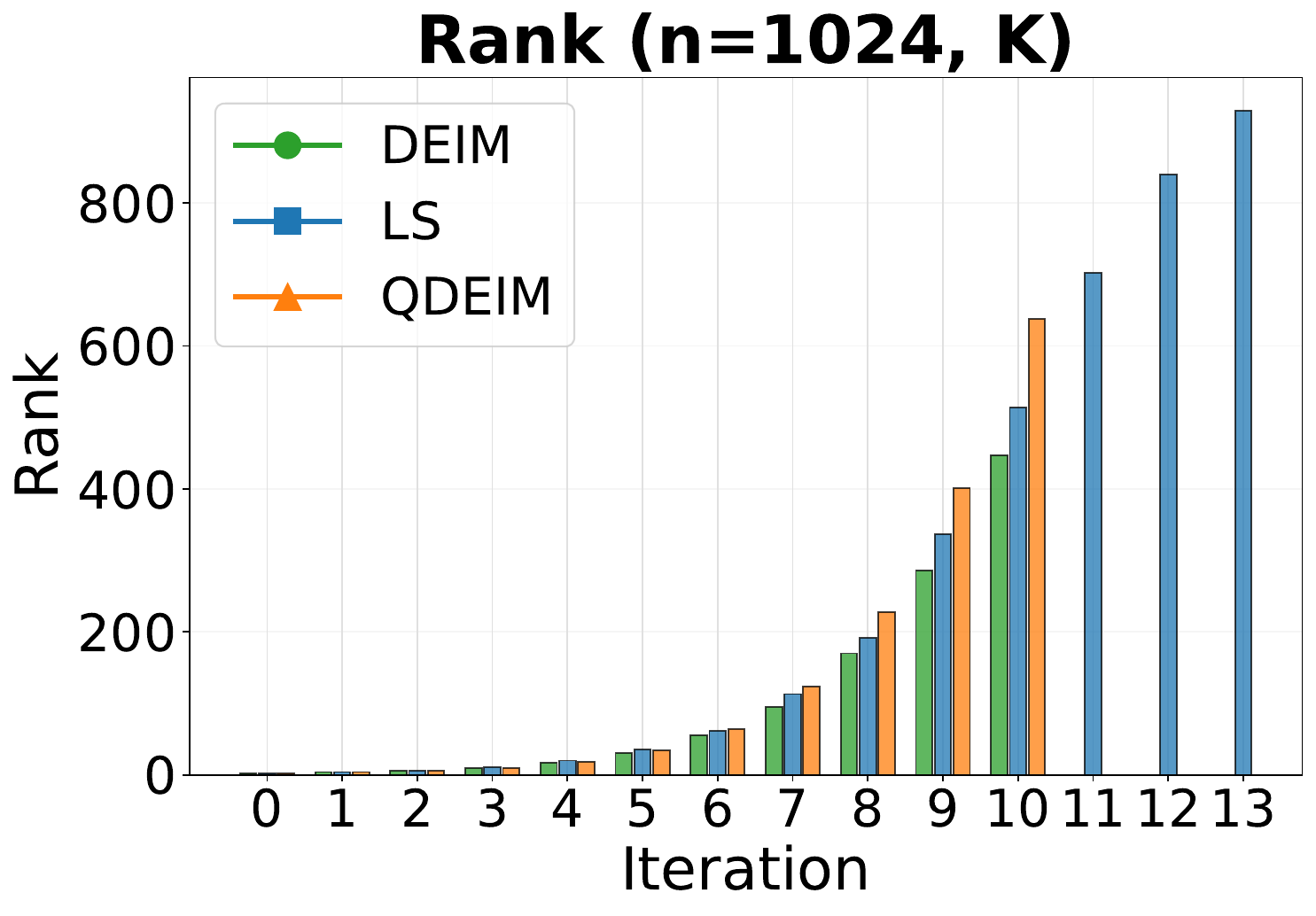}
    \includegraphics[width=0.33\linewidth]{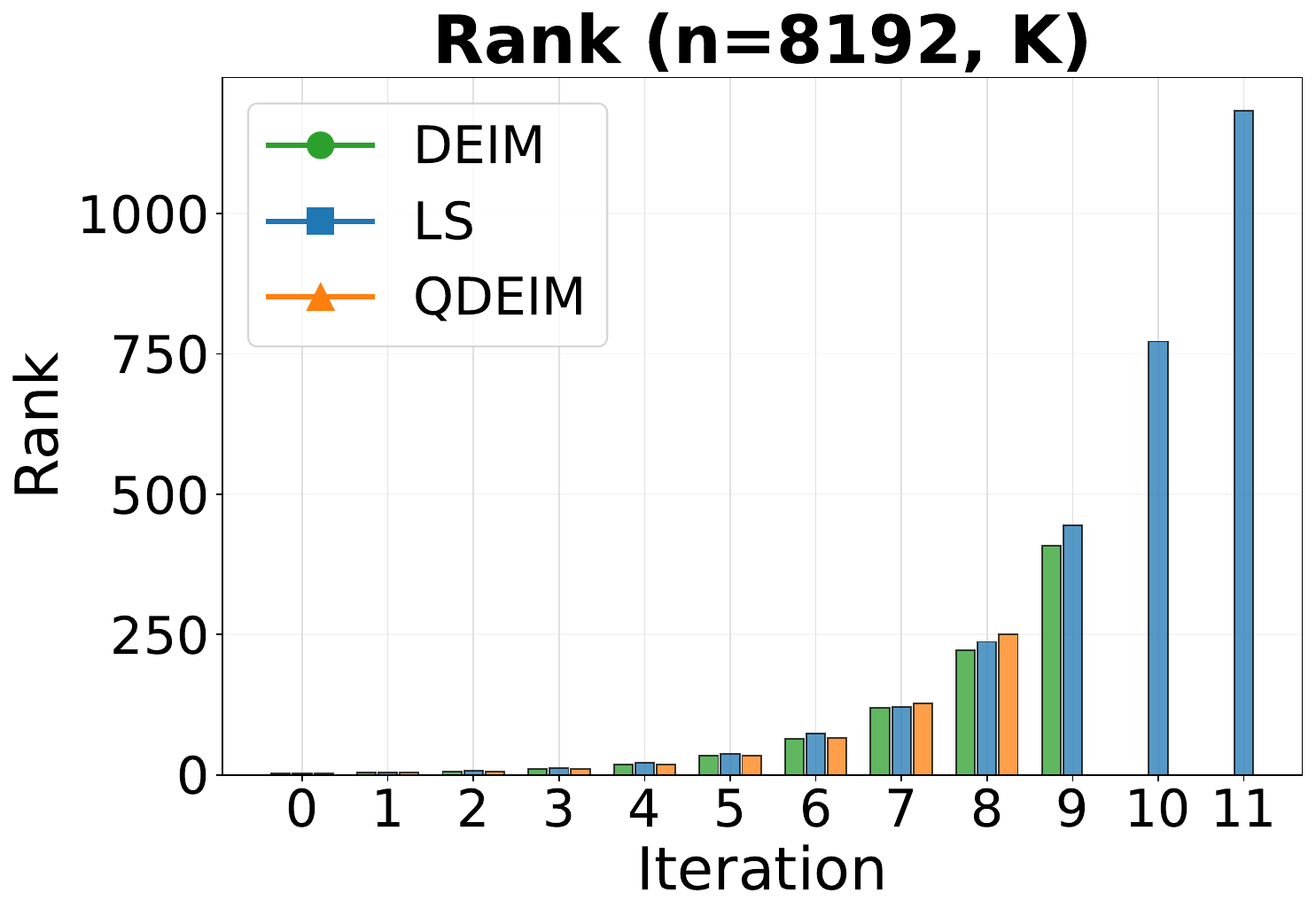}
    \includegraphics[width=0.32\linewidth]{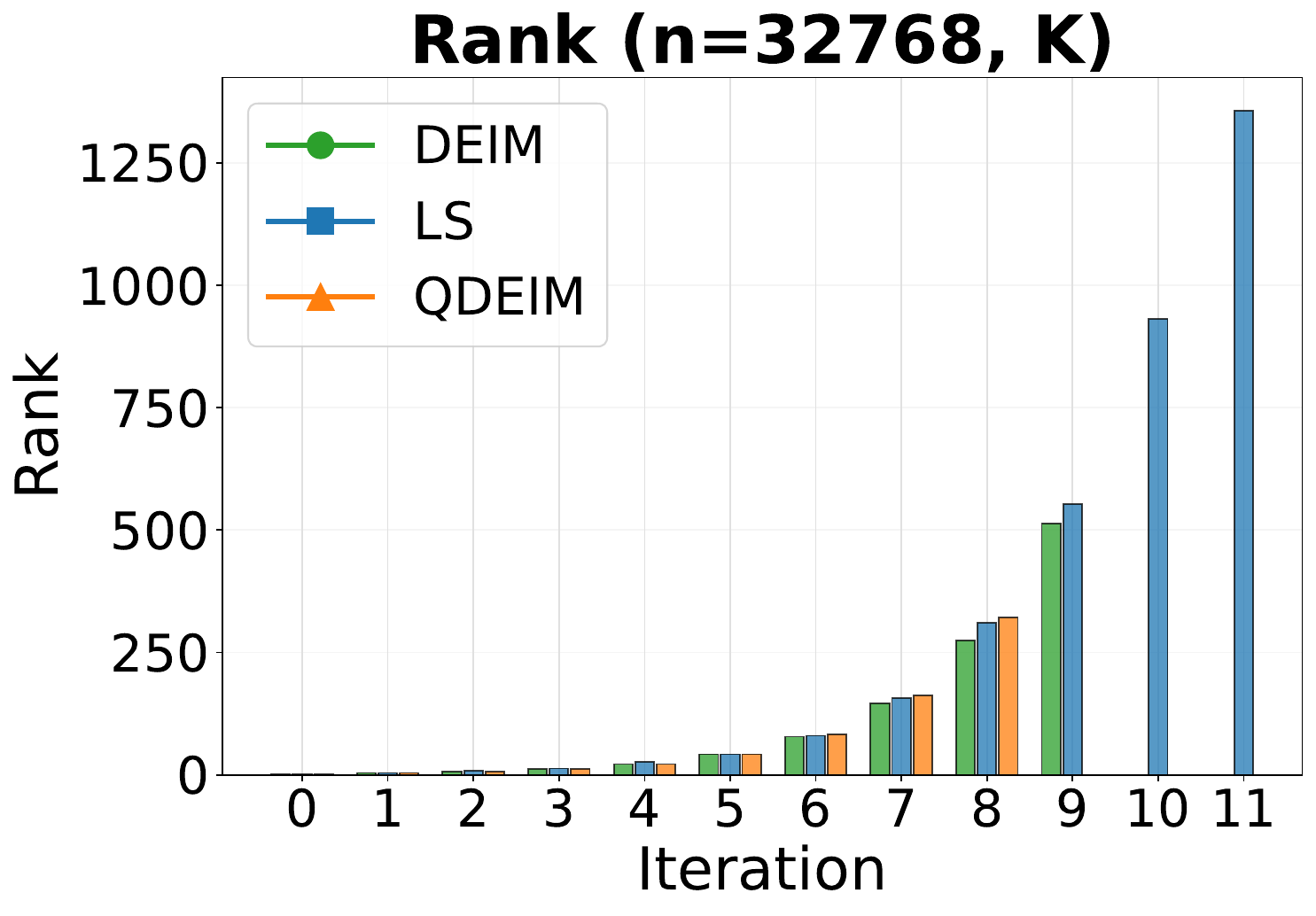}
    \caption{The top row presents the error (solid) and the cumulative time (dashed) for the Cross-DEIM implementation for three sizes of the $K$-matrix. The second row presents the final rank of the solution at each iteration of Cross-DEIM. For this problem the stopping tolerance was set to $10^{-6}$.}
    \label{fig:indexselection2}
\end{figure*}

First the two test matrices are investigated in~\autoref{fig:indexselection} and \autoref{fig:indexselection2} for three sizes using the index selection algorithms considered in the paper, here we present the results from a single run with a fixed seed to show the algorithms specific behaviors. 

For the Hilbert matrix we see fast convergence to the tolerance of $10^{-9}$ in 6-7 Cross-DEIM iterations for the DEIM-variations for all the problem sizes. The quality of the LS selection is clearly worse and for every size more iterations are required. The time for the same number of iterations is always lower for LS compared with DEIM and QDEIM. The final time is lower except for one case which is unusually ill-behaved. Notably the LS method extends the index set less aggressively than DEIM and QDEIM for the Hilbert matrix. The largest case behaves a bit differently in its convergence, and the first few iterations do not make any significant progress in the convergence but all three methods manages to converge at the expected rate in the end. Changing the seed for the random number generator would change the profile of the convergence.

The $K$-matrix on the other hand is more difficult to approximate and the quality of the index selection plays a bigger role. It is notable that the Cross-DEIM algorithm converges in at least one iteration less using QDEIM for all problem sizes although with a larger rank than DEIM and LS for that iteration. 

The index set chosen by LS rarely generates an error that is comparable to DEIM and QDEIM for the same number of Cross-DEIM iterations. Instead further iterations are required and the rank grows quickly; therefore, the final time is worse for LS. It is important to note that the error presented and the stopping criteria is based on the change of the solution between iterations.

\subsection{Performance of the Fast Poisson solver}
\begin{figure}[t]
    \centering
    \includegraphics[width=1\linewidth]{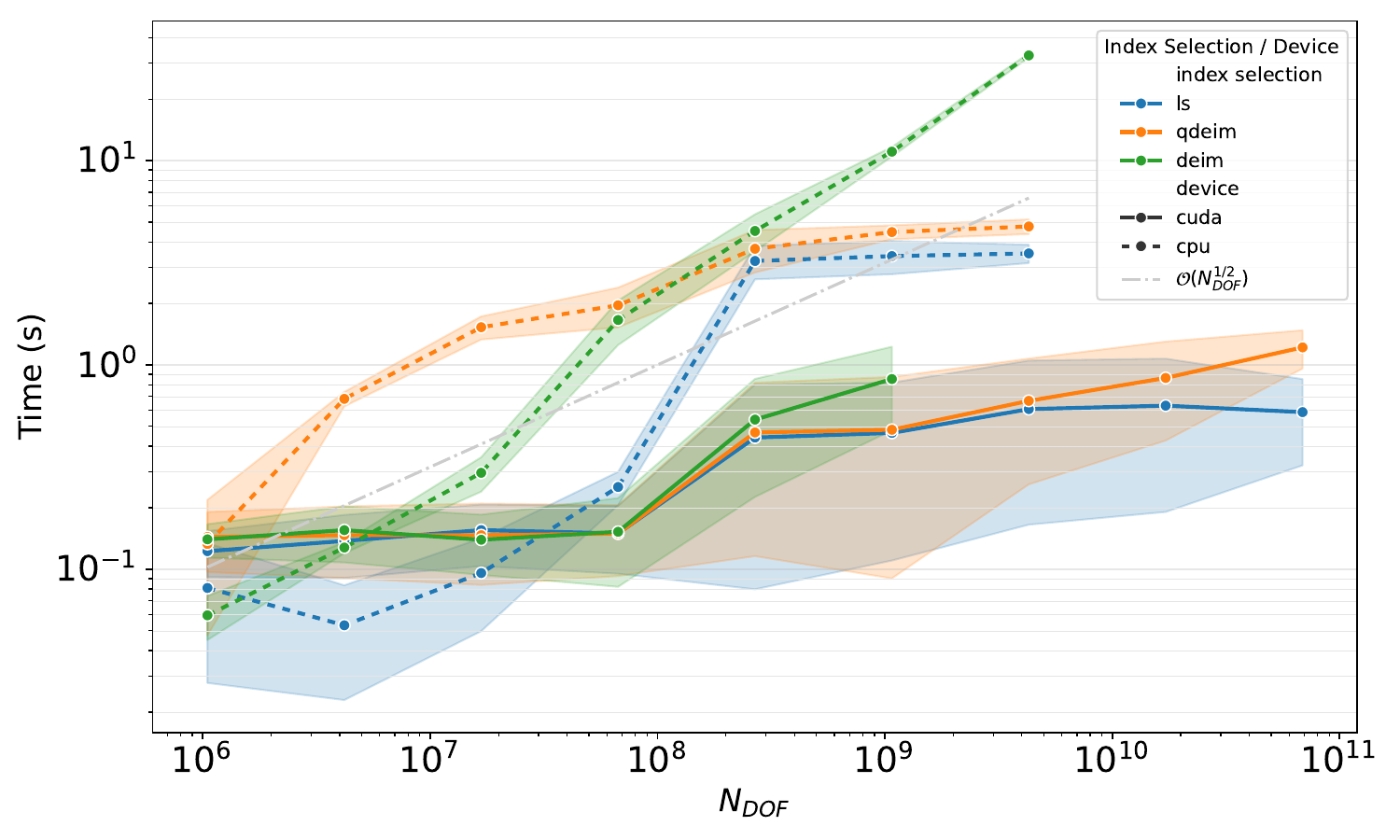}
    \caption{Time here is the wall-clock-time of the whole application. It is run with ten seeds on GPU (solid) and CPU (dashed), the mean and standard deviation is presented. The variation between runs of the same seed is very small.}
    \label{fig:PoissonWCT}
\end{figure}

This section describes the Poisson solvers's performance with respect to the number of grid-points on CPU and GPU and for the different index selection methods. In \autoref{fig:PoissonWCT} the wall-clock-time for the entire application is presented with different problem-sizes.
\begin{figure}[h!]
    \centering
    \includegraphics[width=1\textwidth]{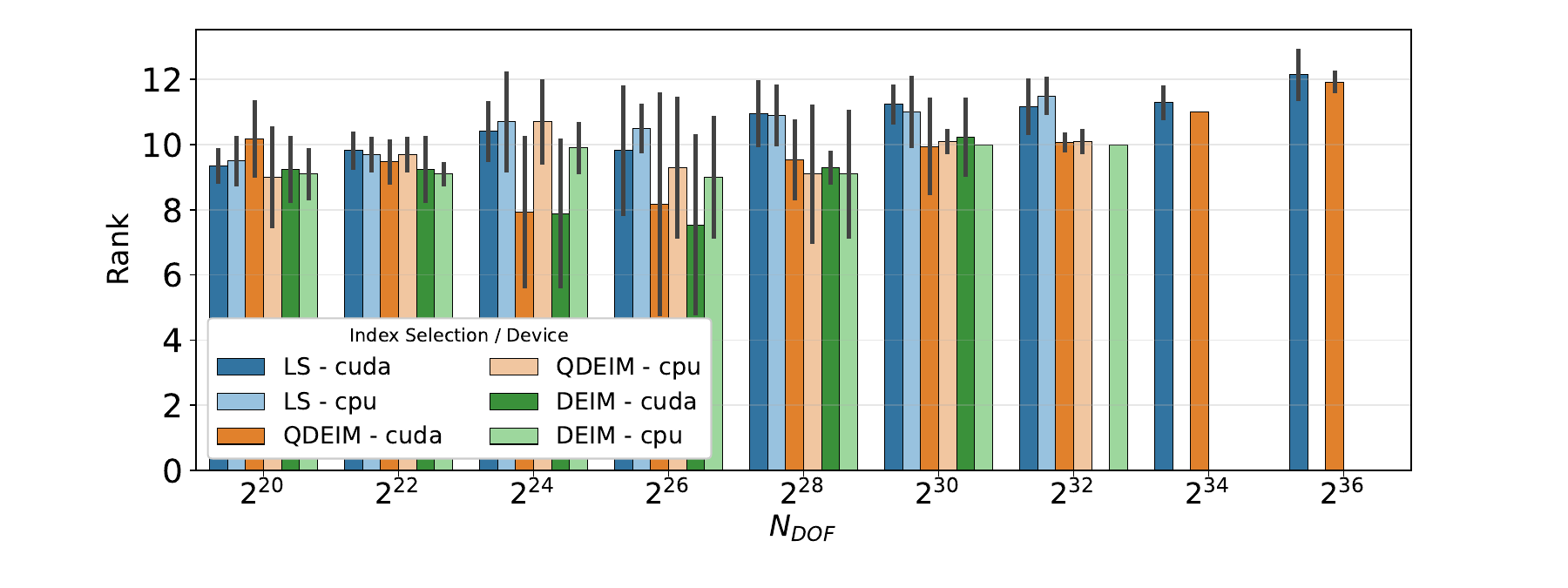}
    \caption{The final Rank of the solution for the different problem sizes solving the Poisson problem. Showing the mean value and standard deviation over 10 seeds. Note: that the same seed in PyTorch does not guarantee the same results on GPU and CPU.}
    \label{fig:rank_poisson}
\end{figure}
In these runs multiple seeds are utilized and the mean time is presented together with the standard deviation. The GPU implementation is clearly preferable for larger problems with a shorter time-to-solution and better scaling with respect to larger problems. Notably the QDEIM and LS selections out-perform DEIM for large problems on both CPU and GPU with a slight edge for LS.

As with the Hilbert matrix the index selection using LS can rival the quality of QDEIM and DEIM. This is seen in \autoref{fig:rank_poisson} which again shows the final rank for the solution depending on the problem size. The rank is found to be similar for the different selection methods although LS generally requires a larger rank to achieve the same accuracy. Note that the final rank does not necessarily describe the computational cost of the algorithm as the intermediate solutions inside the solver can be of a larger rank. 

\begin{figure}
    \centering
    \includegraphics[width=1\textwidth]{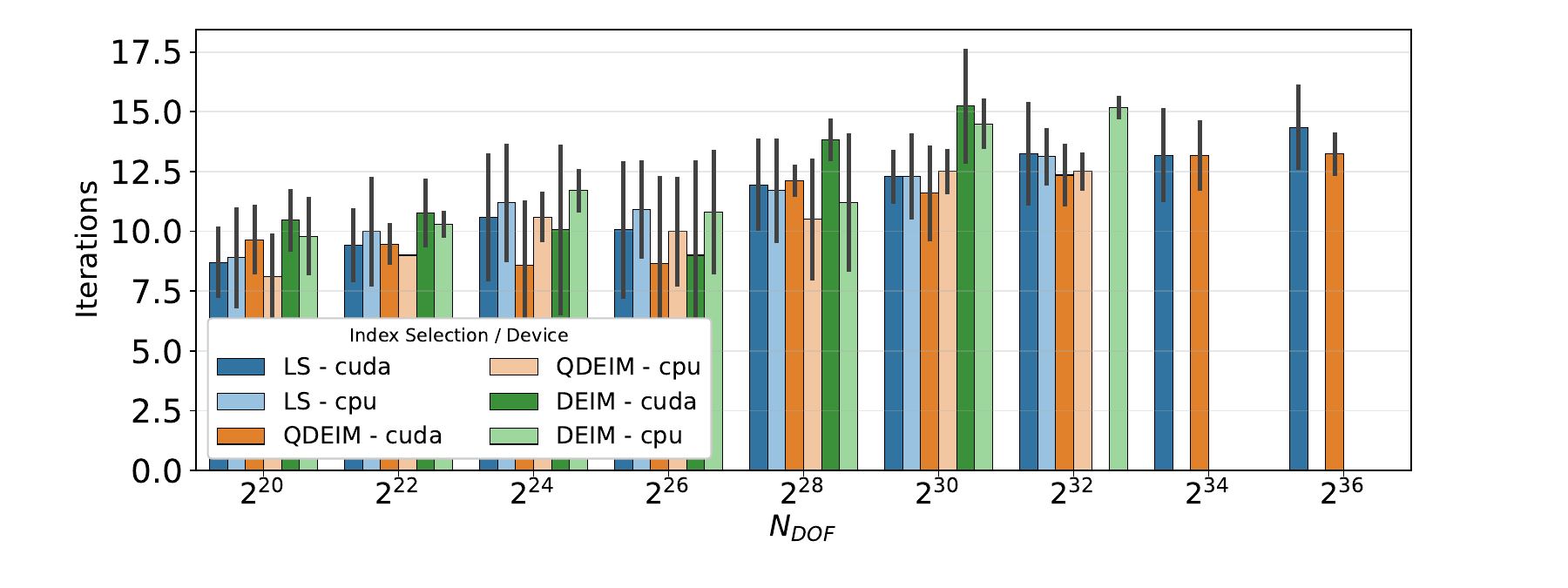}    \caption{Number of Cross-DEIM iteration for the different problem sizes of the Possion problem. Showing the mean value and standard deviation over 10 seeds. Note: that the same seed in PyTorch does not guarantee the same results on GPU and CPU.}
    \label{fig:iter_poisson}
\end{figure}
\begin{figure}[t]
    \centering
    \includegraphics[width=0.5\linewidth]{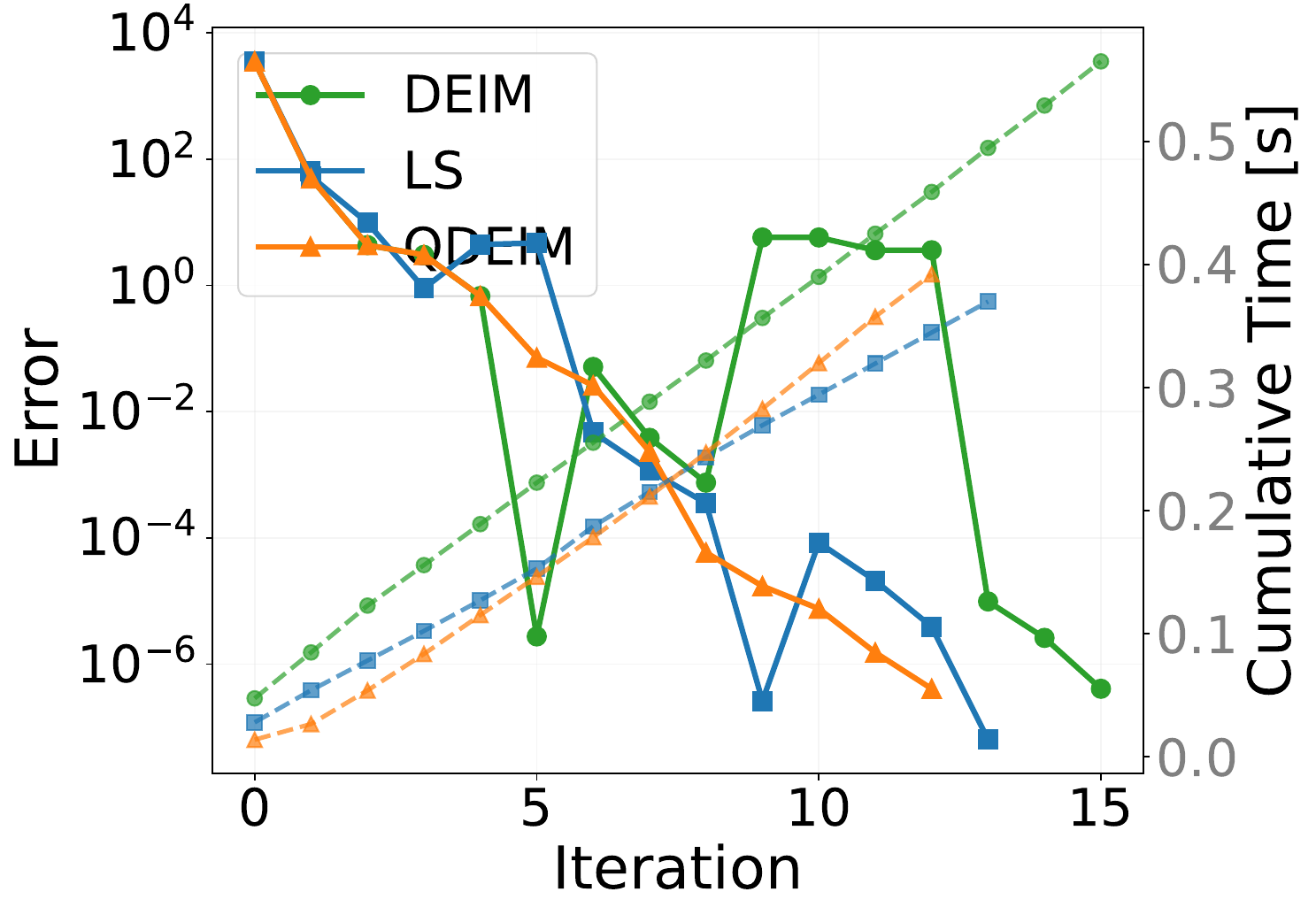}
    \includegraphics[width=0.49\linewidth]{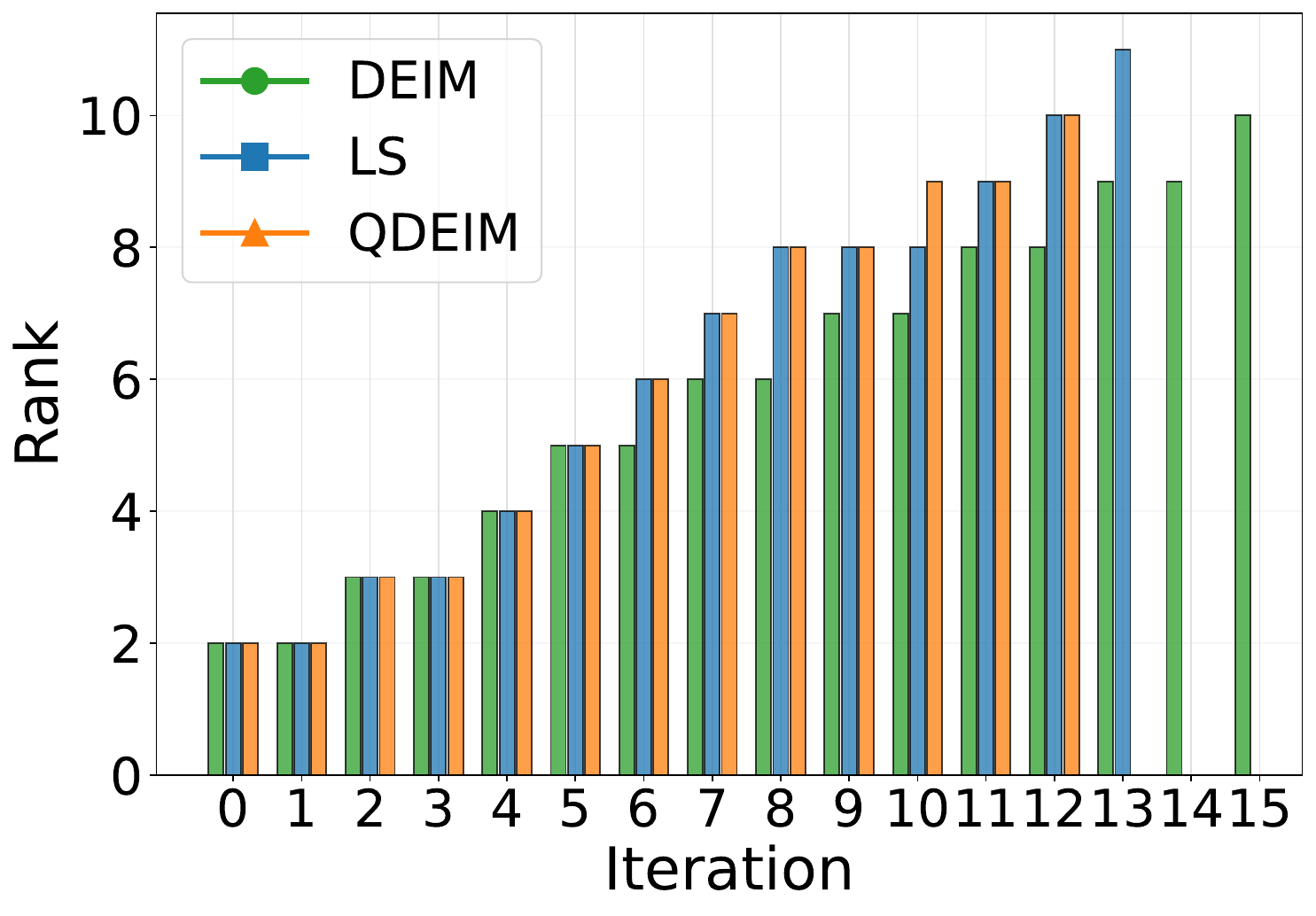}
    \caption{Left: Convergence and cumulative time-cost fort the Cross-DEIM algorithm for the Poisson problem with a fixed size of $\DOF=2^{30}$.}
    \label{fig:possionErrorTiming}
\end{figure}

The number of Cross-DEIM iterations is shown in \autoref{fig:iter_poisson}, for most cases LS manages to find a solution to the same tolerance in a similar number of iterations. DEIM performs worse for larger sizes and quickly requires more iterations than LS and QDEIM. The per iteration time-cost is higher than LS and QDEIM as well even when the rank is lower as shown in~\autoref{fig:possionErrorTiming}. As hypothesized in the background, the cost of an extra iteration can be amortized by the cheaper index selection. We also note that it is hard to compare algorithms with this erratic convergence (DEIM and LS) with a slight change in the tolerance DEIM could have stopped at five iterations and LS at nine. This would have flipped order of methods' performance completely. 

\begin{figure}[h]
    \centering
    \includegraphics[width=1\linewidth]{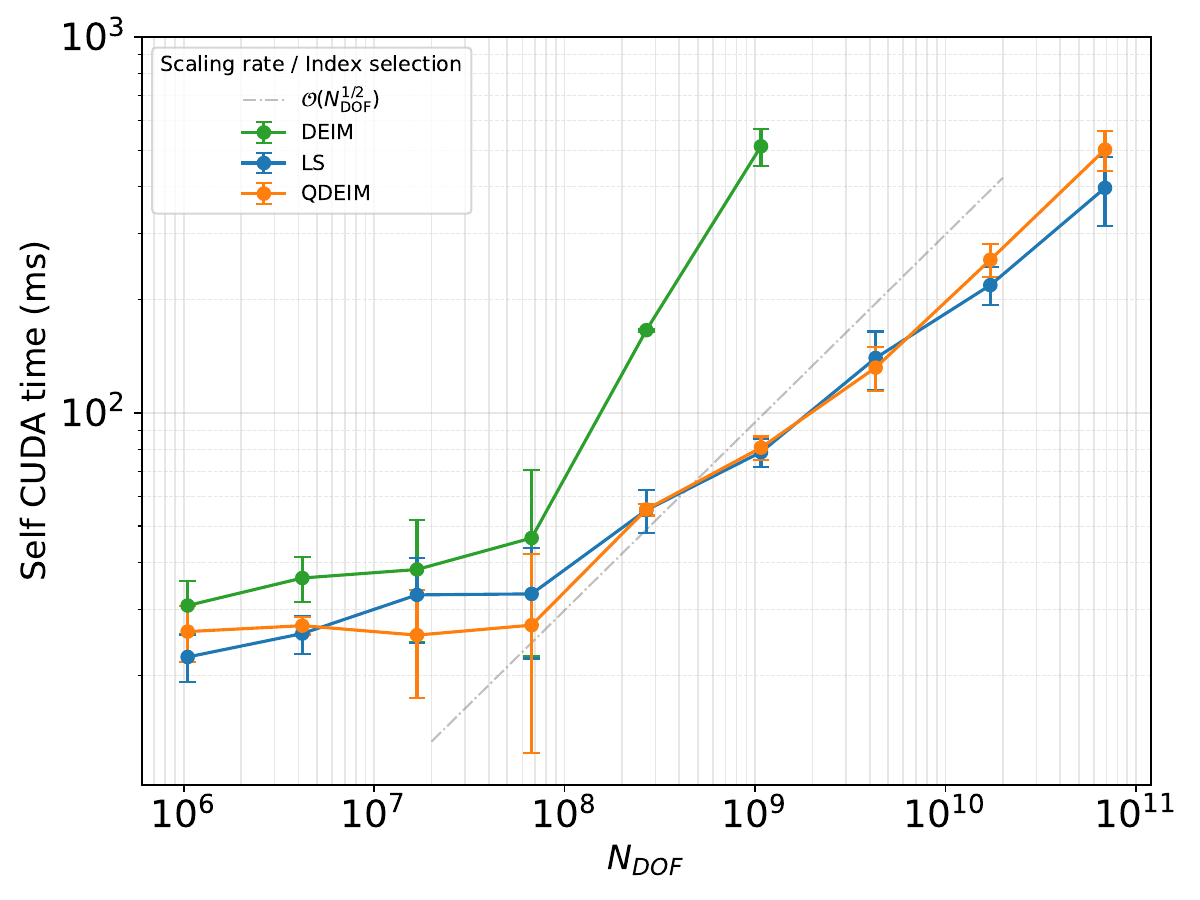}
    \caption{Solving the Poisson Equation for different sizes using a A100 with the fast Cross-DEIM solver using the different selection strategies. Timings gathered with \texttt{torch.profiler}. A clear $\mathcal{O}(\DOF^{1/2})$-behavior can be observed for QDEIM and LS.}
    \label{fig:scalingPoissonSolver}
\end{figure}

Using the timings from the \texttt{torch.profiler} we can show that the expected sublinear cost can be measured for the kernels~\autoref{fig:scalingPoissonSolver} as predicted in \autoref{eq:On12}. The LU factorization in the DEIM calculation is the dominating for Cross-DEIM using DEIM. Therefore, DEIM can quickly be ruled out as unviable compared with QDEIM and LS for larger problems. 
When the problem size grows further, LS performs slightly better than QDEIM both in terms of time-to-solution and the profiled kernel as QDEIM scales worse with the problem size and rank.

\section{Conclusions}
We have adapted a set of low-rank routines useful for cross-approximations used for solving PDEs and investigated a fast transpose-free FFT based Poisson solver which shows sub-linear scaling with respect to the number of grid-points. The code runs on multi-core CPUs as well as on GPUs and enables the study of much larger problems than previously was feasible.

It is clear that there is a role for simple randomized algorithms, especially when the problem size increases but the problem's low-rank is preserved. Cases where the index selection is proved harder for LS, then, a higher rank were generally required to achieve the same error as with QDEIM, but the cost of LS scales much better with the number of unknowns and the rank. 

Note that there are specialized QR implementations that might enable better performance both for the \texttt{scross} algorithm as well as for the index selection. The factorization methods are by far the most expensive components.    


\begin{credits}
\subsubsection{\ackname}
The authors acknowledge Advanced Research Computing at Virginia Tech for providing computational resources and technical support that have contributed to the results reported within this paper. URL: https://arc.vt.edu/. 

DA is supported by the U.S. Department of Energy, Office of Science, Advanced Scientific Computing Research (ASCR), under Award Number DE-SC0025424. This material is based upon work supported, in part, by the National Science Foundation under Grant No DMS-2436319 and Virginia Tech.

This material is based upon work supported by the National Science Foundation under Grant No. DMS-2424139 while the DA were in residence at the Simons Laufer Mathematical Sciences Institute in Berkeley, California, during the Fall 2025 semester.


\end{credits}
%
%
%

\bibliographystyle{splncs04}
\bibliography{ref}
\end{document}